\documentclass[aps,prl,twocolumn,showpacs,10pt]{revtex4-2}
\usepackage{subfiles} 
\usepackage{amssymb,amsmath}
\usepackage{bm, graphicx, amsmath}
\usepackage{amsthm,relsize}
\usepackage{array}
\usepackage{bbm}
\usepackage{comment}
\usepackage{tabularray}
\usepackage{xcolor}
\usepackage[section]{placeins}
\usepackage{mathrsfs}
\usepackage{multirow}
\usepackage[normalem]{ulem}
\usepackage{lipsum}
\usepackage{enumitem}
\usepackage{tikz}
\usetikzlibrary{3d}
\usepackage{changepage}
\usepackage[mathscr]{eucal}
\usepackage{mathrsfs}
\usepackage{color}
\usepackage{hyperref}
\hypersetup{breaklinks=true}
\usepackage{xr-hyper}
\newtheorem{theorem}{Theorem}

\newtheorem{lemma}{Lemma}

\newcommand{\Tr}{\operatorname{Tr}}
\newcommand{\Span}{\operatorname{span}}
\newcommand{\Range}{\operatorname{range}}

\newcommand{\bra}[1]{\langle #1|}
\newcommand{\ket}[1]{|#1\rangle}
\newcommand{\braket}[2]{\langle #1|#2\rangle}
\newcommand{\ketbra}[2]{\ket{#1}\bra{#2}}

\definecolor{cadmiumgreen}{rgb}{0.0, 0.42, 0.24}

\begin{document}


\title{Quantum state exclusion for group-generated ensembles of pure states}
\author{A. Diebra$^{1}$, S. Llorens$^{1}$,  E. Bagan$^{1}$, G. Sent\'{i}s$^{1,2}$, and R. Mu\~noz-Tapia$^{1,3}$}

\affiliation{$^{1}$F\'{i}sica Te\`{o}rica: Informaci\'{o} i Fen\`{o}mens Qu\`antics, Universitat Aut\`{o}noma de Barcelona, 08193 Bellaterra (Barcelona), Spain
}
\affiliation{$^{2}$Ideaded, Carrer de la Tecnologia, 35, 08840 Viladecans, Barcelona, Spain
}
\affiliation{$^{3}$H. H. Wills Physics Laboratory, University of Bristol, Tyndall Avenue, Bristol, BS8 1TL, United Kingdom
}

\begin{abstract}  
Quantum state exclusion is the task of determining which states from a given set a system was not prepared in.
We provide a complete solution to optimal quantum state exclusion for arbitrary sets of pure states generated by finite groups, establishing necessary and sufficient conditions for perfect (zero-error conclusive) exclusion. When perfect exclusion is impossible, we introduce two natural extensions: minimum-error and unambiguous exclusion. For both, we derive the optimal protocols and present analytical expressions for the corresponding failure probabilities and measurements, providing additional insight into how quantum states encode information.
\end{abstract}  

\maketitle
\textit{Introduction}.
In both scientific and everyday contexts, the ability to confidently exclude what did not occur, based on partial information, is often more efficient than attempting to determine what actually did. This principle is widely applied in classical scenarios, where exclusion serves as a powerful tool for narrowing possibilities and guiding decision-making. For instance, in medical diagnostics, ruling out potential diseases is crucial for refining treatment options, even when the exact disease remains unidentified~\cite{mcgee2012evidence}. Similarly, in engineering system testing, excluding certain failure modes is often more advantageous than trying to pinpoint the exact faulty process~\cite{birolini2017reliability}. Additionally, in network security, excluding potential vulnerabilities is usually more efficient than attempting to track down the precise nature of a threat~\cite{stallings2020network}.

The very same principle extends into the quantum realm through a task named quantum state exclusion (QSE)~\cite{band_2014}. Given a quantum system prepared in an unknown state drawn from a predetermined set, the goal is to exclude that the state of the given system is a particular one from that set. This task, also known as quantum antidistinguishability~\cite{heinosaari_2018}, is particularly relevant in the context of quantum state assignments, where the compatibility of different state assignments must be assessed~\cite{brun_2002,caves_2002}. It also plays a crucial role in the debate surrounding the physical reality of quantum systems~\cite{pusey_2012}. 

In contrast to the traditional approach of quantum state discrimination~\cite{chefles_2000,Bergou_2010,Bae_2015}, where the goal is to identify the exact state of a system, QSE focuses on ruling out certain states, offering a different perspective on the information that can be extracted from quantum systems. 
QSE plays a role in quantum resource theories: Any quantum resource can be quantitatively linked to the relative advantage it offers in a state exclusion task compared to free resources~\cite{ducuara_2020,uola_2020}. QSE has also been shown to provide~an operational interpretation of the Choi rank of a quantum channel~\cite{stratton_2024}. This rank, among other properties, places bounds on the minimum number of Kraus operators required to decompose a channel. The role of contextuality in QSE has been discussed in Ref.~\cite{context_2024}. Conversely, non-contextual inequalities have been derived from QSE~\cite{leifer_2020}, with implications for communication complexity~\cite{havlivcek_2020,heinosaari_2024}. More recently, Chernoff error exponents for QSE have been calculated in Ref.~\cite{mishra_2024}. Additionally, in quantum state discrimination with multiple copies, it has been shown that the optimal procedure may, in some cases, involve excluding specific states~\cite{sentis2022}. QSE has also been demonstrated experimentally with single photons in Ref.~\cite{PhysRevResearch.5.023094}.

Just as in quantum state discrimination, finding optimal protocols for QSE in a fully general setting remains a challenging problem. Beyond the binary case (also known as hypothesis testing), multihypothesis quantum state discrimination has only been solved when a certain degree of symmetry can be invoked~\cite{sasaki_1999,barnett_2001,eldar_2001,eldar_2004,nakahira_2012,nakahira_2013,sentis_2016,sentis_2017,llorens-2024,skotiniotis_2024}. This same symmetry principle has been applied in the context of QSE, where it has been used to derive tight bounds for circulant sets of pure quantum states~\cite{johnston_2023}.

As far as we know, all existing literature on QSE has focused on perfect exclusion, that is, protocols that enable the error-free conclusive exclusion of one or more states. However, perfect QSE is only possible within a limited region of the parameter space defining the set of states (or hypotheses), with necessary and/or sufficient conditions delineating this region already established. Moreover, many existing results are non-constructive: While they confirm the existence of an error-free exclusion measurement, they do not provide its explicit form~\cite{johnston_2023}.

 In this Letter, we present a complete analytical solution to optimal QSE for group-generated sets of pure states, revealing additional facets of the problem. Specifically, we derive the minimum error probability of exclusion for any finite group and across the entire parameter space, both within and outside the perfect exclusion region, providing a specific measurement that achieves this minimum. We also prove that this optimal measurement can always be taken to be sharp. Our approach relies on the Gram matrix~\cite{horn_2012} formalism,  exploits the duality properties of semidefinite programming~\cite{boyd_2004,watrous_2018} and makes extensive use of group representation theory~\cite{hamermesh1989group,steinberg2011representation,christandl2006structurebipartitequantumstates}. 
 
We also investigate the zero-error (unambiguous) approach~\cite{crickmore_2020}, which allows for inconclusive answers while forbidding errors, even outside the perfect QSE region. In this context, we provide an analytical expression for the probability of  obtaining such inconclusive answers for any group-generated set of pure states. 

We show that the minimum probabilities of error and inconclusive answers are both determined by the eigenvalues of the Gram matrix.
The same formulas seem to provide a useful upper bound for these probabilities even in general, nonsymmetric cases. This is in agreement with intuition, as symmetry inherently increases indistinguishability and reduces the ability to exclude specific objects, making discrimination and exclusion more challenging.

This Letter is organized as follows. We first present the formulation and structure of the problem, followed by our main results and an outline of the proof of a central lemma, which establishes that the optimal measurements can be constructed from specific rank-one operators. We conclude with a summary of our findings.

\noindent\textit{Formulation and structure of the problem}. This Letter addresses the task of excluding any single preparation from a given ensemble \mbox{$\mathcal{E} = \{\eta_i, \rho_i\}_i$} of possible states, where \mbox{$\rho_i = |\psi_i\rangle\langle\psi_i|$} occurs with prior probability $\eta_i$. The exclusion is performed through a quantum measurement, represented by a positive operator-valued measure (POVM) $\{\Pi_j\}_j$ on the Hilbert space $\cal H$ of the states, with each outcome~$j$ indicating that the state $\rho_j$ is ruled out.  
For certain ensembles of states perfect exclusion is possible, namely, there exists a POVM such that $\Tr(\Pi_i\rho_i)=0$ for all $i$. However, as explained in the introduction, we also consider ensembles for which this does not necessarily hold. In analogy with state discrimination, we examine two natural extensions of perfect QSE: minimum-error and unambiguous exclusion.

Minimum-error exclusion aims to minimize the average probability of incorrect exclusion, denoted by $P$, over the set of all possible POVMs. This optimization can be formulated as a semidefinite program (SDP):
\begin{multline}
   P^{\rm min}=\min_{\{\Pi_i\}} \sum_{i}\eta_i\Tr(\Pi_i\rho_i),\\[-.3em]
    \mbox{subject to}\  \sum_i\Pi_i = \openone, \quad \Pi_i\geq 0\ \mbox{for all  $i$}.
    \label{eq:min_error_sdp}
\end{multline}

Alternatively, unambiguous exclusion imposes the strict constraint that no errors occur, typically requiring an additional POVM element, $\Pi_Q$, associated with an inconclusive outcome. When this outcome is obtained, no definitive decision can be made, so the optimal protocol seeks to minimize its average probability $Q$. This too can be formulated as an SDP:
\begin{multline}\label{eq:zero_error_sdp}
    Q^{\rm min}=\min_{\{\Pi_i\}} \sum_i\eta_i\Tr(\Pi_Q\,\rho_i),\\[-.3em]
    \mbox{subject to}\ \Pi_Q+ \sum_i\Pi_i = \openone,\quad \Pi_Q\geq 0, \\[-.2em]
   \Pi_i\geq 0, \quad \Tr(\Pi_i\rho_i)= 0 \ \mbox{for all $i$}.
\end{multline}
Perfect exclusion corresponds to~\mbox{$P^{\rm min}=Q^{\rm min}=0$}. We refer to $P$ and $Q$ as failure probabilities.

In this work, we focus on quantum ensembles that are group-generated,  
${\cal E}\!=\!\{\ket{\psi_g}\! =\! U_g \ket{\psi}\,|\, {g\in{\cal G}}\}$,  
where each state in the ensemble is obtained by applying a finite-dimensional unitary (linear or projective) representation,  
\mbox{$U: g \mapsto U_g \in {\cal U}({\cal H})$},  
of a finite group~$\cal G$ to a seed state~$\ket{\psi}$. Furthermore, we assume that all states are equally probable, i.e., $\eta_g = 1/|\mathcal{G}|$, for all $g\in{\cal G}$, where $|\mathcal{G}|$ is the order of $\mathcal{G}$.  

Any such representation can be decomposed into~a~direct sum of irreducible representations \mbox{(irreps for short)},  
\mbox{$U_g = \bigoplus_{\mu}U_g^\mu\otimes\openone_{m_\mu}$},  
where $\openone_d$ stands for the $d$-di\-men\-sional identity operator, $\mu$ labels the distinct irreps, and $m_\mu$ denotes their multiplicity. Accordingly, the Hilbert space~$\cal H$ breaks into orthogonal subspaces~\cite{hamermesh1989group,steinberg2011representation,christandl2006structurebipartitequantumstates} 
\mbox{as $ \mathcal{H} = \bigoplus_{\mu}\mathcal{H}_\mu\otimes\mathbb{C}^{m_\mu}$,}
where $\mathcal{H}_\mu$ carries the irrep $\mu$, while $\mathbb{C}^{m_\mu}$ denotes the multiplicity space, over which~$U$ acts trivially.  
This decomposition allows us to write the seed state as \mbox{$\ket{\psi}=1/\sqrt{|\mathcal{G}|}\sum_{\mu}\sqrt{d_\mu}\ket{\psi_\mu}$},
where $d_\mu$ is the dimension of the irrep $\mu$ and \mbox{$\ket{\psi_\mu}\in\mathcal{H}_\mu\otimes\mathbb{C}^{m_\mu}$} are non-normalized bipartite states. The factors $1/\sqrt{|{\cal G}|}$ and$\sqrt{d_\mu}$ have been introduced for later convenience. 
The states~$|\psi_\mu\rangle$ can be written in a Schmidt form:
\begin{equation}
    \ket{\psi_\mu} = \sum_{k=1}^{r_\mu}\sqrt{\alpha_{k}^{\mu}}\,|v_k^\mu\rangle|u_k^\mu\rangle ,
    \label{bipartite_decomp}
\end{equation}
where \mbox{$r_\mu = \min\{d_\mu,m_\mu\}$}, $\alpha_k^\mu\geq0$, for all $\mu$ and $k$, and both $\{\ket{v_k^{\mu}}\}_{k=1}^{d_\mu}$ and $\{\ket{u_k^\mu}\}_{k=1}^{m_\mu}$ are orthonormal bases of~$\mathcal{H}_\mu$ and~$\mathbb{C}^{m_\mu}$ respectively. Using the great orthogonality theorem~\cite{hamermesh1989group,steinberg2011representation,christandl2006structurebipartitequantumstates,supplemental} , one can show that the operator $\Omega = \sum_{g}U_g\ketbra{\psi}{\psi}U_g^\dagger$, which represents the (unnormalized) density matrix of the ensemble (ensemble operator for short), is diagonal in this basis:  
\begin{equation}\label{Omega}
    \Omega = \bigoplus_\mu \openone_{d_\mu} \otimes \sum_{k=1}^{r_\mu} \alpha_k^{\mu} \ketbra{u_{k}^\mu}{u_{k}^\mu}.
\end{equation}

As in the case of quantum state discrimination, all the information required for exclusion is encoded in the Gram matrix $G$ of the ensemble~\cite{supplemental}, whose entries are the overlaps between the states: $G_{g,h} = \braket{\psi_g}{\psi_h}$, with $g, h \in \mathcal{G}$. The symmetry of the problem —specifically, the representation $U$ that generates the ensemble— determines the structure of the Gram matrix, which belongs to the algebra of the (right-)regular representation of the group~\cite{hamermesh1989group}. Explicit examples 
are provided in the Supplemental Material~\cite{supplemental}.  

For any ensemble $\mathcal{E}$, the columns of the square root of the Gram matrix, $S = \sqrt{G}$, define an associated ensemble~$\mathcal{E}_S$ from a $|\cal G|$-dimensional Hilbert space ${\cal H}_S$, which shares the same Gram matrix, $G$. If~$\mathcal{E}$ is generated by a group $\mathcal{G}$, so is $\mathcal{E}_S$, with their states belonging to the (left-)regular representation of $\mathcal{G}$~\cite{supplemental}.

The Gram matrix \( G \) and the ensemble operator \( \Omega \) are closely related. Specifically, among the eigenvalues of \( G \) —denoted as \( \lambda_a \), with \( \lambda_1 \geq \lambda_2 \geq \dots \geq \lambda_{|\mathcal{G}|} \)~— those that are strictly positive coincide with the non-zero eigenvalues of \( \Omega \)~\cite{supplemental}, which can be read off from Eq.~(\ref{Omega}).

\noindent\textit{Results.}  
For ensembles of pure states generated by a finite group $\cal G$, the minimum error probability for excluding a state, as defined in Eq.~(\ref{eq:min_error_sdp}), is determined by the eigenvalues $\{\lambda_a\}_{a=1}^{|{\cal G}|}$ of the Gram matrix:  
\begin{equation}
\label{the Pe}
P^{\rm min} = \Bigg[ \frac{1}{|\mathcal{G}|} \max\!\Bigg( 0, \sqrt{\lambda_1} - \sum_{a > 1} \sqrt{\lambda_a} \Bigg) \Bigg]^2.
\end{equation}  
Similarly, for unambiguous QSE, the probability of the inconclusive outcome is given by 
\begin{equation}
\label{eq:Pe-unambiguous} 
Q^{\rm min} = \frac{\Tr(\sqrt{G})}{|\mathcal{G}|} \max\Bigg( 0, \sqrt{\lambda_1} - \sum_{a > 1} \sqrt{\lambda_a} \Bigg).
\end{equation}   
These results establish that the condition  
\begin{equation}
\label{the condition}
\sqrt{\lambda_1} \leq \sum_{a> 1} \sqrt{\lambda_a}  
\end{equation}  
is both necessary and sufficient for perfect QSE to be possible in group-generated ensembles. As a special case, Theorem 5.1 in Ref. \cite{johnston_2023} —which applies to circulant matrices— follows directly from this result, now extended to arbitrary finite groups.

Equations~(\ref{the Pe}) and~(\ref{eq:Pe-unambiguous}) follow from the lemma below. This lemma also provides the optimal measurements for the two approaches considered in this Letter. Both of them are represented by rank-1 (group-)covariant POVMs. 

Note that if the largest eigenvalue \(\lambda_1\) has multiplicity greater than one, condition~(\ref{the condition}) is automatically satisfied, ensuring perfect QSE. This occurs whenever the largest Schmidt coefficient (or eigenvalue of \(\Omega\)), \(\alpha_k^\mu\), corresponds to an irrep \(\mu\) of dimension greater than one.  

The results above extend to unitary projective representations, where \( U_g U_h = e^{i\theta(g,h)} U_{gh} \) for all \( g,h \in \mathcal{G} \). Such representations are not merely of academic interest: spin-\(\frac{1}{2}\) particles transform under a projective representation of the 3-dimensional rotation group. For instance, the ensemble~\( \{|\boldsymbol{n}_\alpha\rangle\}_\alpha \), where \( \alpha \) indexes the vertices of a regular tetrahedron in the Bloch sphere, can be viewed as generated by the Pauli matrices $\{\openone_2,\sigma_x,\sigma_y,\sigma_z\}$, which form a nontrivial projective representation of \( \mathbb{Z}_2 \times \mathbb{Z}_2 \)~\cite{supplemental}. These states also define a symmetric informationally complete POVM~\cite{sic-povm} for qubits.

When the phases \( \theta(g,h) \) are non-trivial, i.e., they cannot be absorbed into the matrices \(\{U_g\}_{g\in\mathcal{G}}\) themselves, the corresponding irreps necessarily have dimension 2 or higher~\cite{hamermesh1989group,steinberg2011representation,christandl2006structurebipartitequantumstates,CHENG2015230}. Thus, by the previous argument, perfect exclusion is always possible for sets generated by such non-trivial projective representations~\cite{supplemental}.

\begin{lemma}\label{lemma:rank-1}
The optimal POVM for QSE can always be chosen to be covariant with a \mbox{rank-1} seed. Specifically, it takes the form  
\mbox{$\{\Pi_g = U_g \ketbra{\omega}{\omega} U_g^\dagger \mid g \in \mathcal{G} \}$} for minimum-error QSE, with the additional POVM element
\mbox{$\Pi_Q=\mbox{\rm \openone}-\sum_{g\in{\cal G}} \Pi_g$} for unambiguous QSE.
Moreover, \mbox{if $\sqrt{\lambda_1} > \sum_{a>1} \sqrt{\lambda_a}$}, the optimal seed state is  
\begin{equation}
\label{eq:non_perf_ansatz}
\ket{\omega} = \frac{1}{\sqrt{|\mathcal{G}|}}\Bigg(\gamma\,|v^1_1\rangle|u^1_1\rangle - \sum_{\mu\neq 1} \sqrt{d_\mu} \sum_{k=1}^{r_\mu} |v_k^\mu\rangle |u_k^\mu\rangle \Bigg)\!,
\end{equation}
where the eigenstate $\ket{v^1_1}\ket{u^1_1}$ corresponds to the largest eigenvalue of $\Omega$ in Eq.~(\ref{Omega}), with $\gamma=1$ for minimum-error QSE, and
\begin{equation}
\label{gamma unamb}
\gamma = \sum_{\mu\not= 1}\sum_{k=1}^{r_\mu}d_\mu{\sqrt{\alpha_k^\mu}\over \sqrt{\alpha^1_1}}
\end{equation}
for unambiguous QSE.
\end{lemma}

\begin{proof}
Consider a POVM of the specified form for some $\ket{\omega}\in\mathcal{H}$. It must hold that $\Phi:=\sum_{g\in\mathcal{G}}U_g\ketbra{\omega}{\omega}U_g^\dag = \openone_{\psi}$, where $\openone_\psi$ is the projector onto the span of the orbit of the seed state $|\psi\rangle$. 
Therefore, applying (the analog of) Eqs.~\eqref{bipartite_decomp} and~\eqref{Omega} to $\Phi$, we see that the state $\ket{\omega}$ must be
\begin{equation}
    \ket{\omega} = \frac{1}{\sqrt{|\mathcal{G}|}}\sum_{\mu}\sqrt{d_\mu}\sum_{j=1}^{r_\mu}|e_j^\mu\rangle|f_j^\mu\rangle,
   \label{eq:POVM_seed} 
\end{equation}
where $\{\ket{e_j^{\mu}}\}_{j=1}^{d_\mu}$ and $\{\ket{f_j^\mu}\}_{j=1}^{m_\mu}$ are orthonormal bases of~$\mathcal{H}_\mu$ and $\mathbb{C}^{m_\mu}$, respectively, with the extra condition that $\text{span}\{\ket{f_j^\mu}\}_j = \text{span} \{\ket{u_k^\mu}\}_k$ for all $\mu$.
The average error probability becomes
\begin{multline}\label{the Pe ve uf}
    P=  |\braket{\omega}{\psi}|^2  =   \\  \frac{1}{|\mathcal{G}|^2}\!\Bigg|\sum_{\mu}d_\mu\!\sum_{j,k=1}^{r_\mu}\sqrt{\alpha_j^\mu}\left|\braket{v_j^\mu}{e_k^\mu}\braket{u_j^\mu}{f_k^\mu}\right|e^{i\theta_{j,k}^\mu}\Bigg|^2,
\end{multline}
where we have written the overlaps in polar form.

To minimize the error probability, we can replace the original ensemble $\cal E$ by its associated ensemble ${\cal E}_S$ and, hence, without loss of generality, take $U$ to be the regular representation.
In the decomposition of this representation, all the irreps of~$\cal G$ appear, each with multiplicity equal to its dimension~\cite{hamermesh1989group,steinberg2011representation,christandl2006structurebipartitequantumstates}, i.e., \mbox{$d_\mu = m_\mu=r_\mu$}. 
Thus, we can always choose mutually unbiased bases $\{\ket{e_j^\mu}\}_j$ and~$\{\ket{v_j^\mu}\}_j$, as well as $\{\ket{f_j^\mu}\}_j$ and $\{\ket{u_j^\mu}\}_j$~\cite{durt_2010}, for the spaces ${\cal H}_{\mu}$ and ${\mathbb C}^{m_\mu}$, respectively, such that
\begin{equation}
    |\braket{e_j^\mu}{v_k^\mu}| = |\braket{f_j^\mu}{u_k^\mu}| = \frac{1}{\sqrt{d_\mu}},\quad \mbox{$1\le j,k\le d_\mu$.}
\end{equation}
With this choice, the average error probability can be expressed as
\begin{equation}
    P = \frac{1}{|\mathcal{G}|^2}\Bigg|\sum_\mu\sum_{j,k=1}^{d_\mu}\sqrt{\alpha_j^\mu}e^{i\theta_{k,j}^\mu}\Bigg|^2.
\end{equation}
Since each eigenvalue $\alpha_k^\mu$ appears in $\Omega$ with multiplicity~$d_\mu$ [see Eq.~\eqref{Omega}], we can rewrite the error probability~as
\begin{equation}\label{sum_eigenvalues}
    P = \frac{1}{|\mathcal{G}|^2}\Bigg|\sum_{a=1}^{|\mathcal{G}|}\sqrt{\lambda_a}e^{i\theta_a}\Bigg|^2 .
\end{equation}

If $\sqrt{\lambda_1}\leq\sum_{a>1}\sqrt{\lambda_a}$, the phases in Eq.~(\ref{sum_eigenvalues}) can be chosen such that the error probability vanishes. This implies that $\Tr(\Pi_g\rho_g)=0$ for all $g \in \mathcal{G}$, and therefore, the probability of obtaining an inconclusive outcome also vanishes. This completes the proof of Lemma \ref{lemma:rank-1} when the above condition holds. 
Note that the optimal seed is infinitely degenerate, as there are infinitely many phase choices that yield $P^{\rm min} = Q^{\rm min} = 0$, corresponding to perfect exclusion.

In contrast, when \(\sqrt{\lambda_1} > \sum_{a>1} \sqrt{\lambda_a}\), the error probability \(P\) in Eq.~(\ref{sum_eigenvalues}) does not vanish. Its minimum occurs when \(\theta_a = \theta_1 + \pi\) for all \(a > 1\). In this case, we adopt the ansatz from Eq.~(\ref{eq:POVM_seed}), with \(\ket{e^\mu_j} \propto \ket{u^\mu_j}\) and \(\ket{f^\mu_j} \propto \ket{v^\mu_j}\), replacing the unbiased bases used previously. Under this choice of bases and phases, Eq.~(\ref{eq:POVM_seed}) gives Eq.~(\ref{eq:non_perf_ansatz}) with \(\gamma = 1\), and Eq.~(\ref{the Pe ve uf}) simplifies to the right-hand side of Eq.~(\ref{the Pe}), denoted as \(P^{\rm primal}\).

The optimality of this ansatz follows from the dual formulation of Eq.~\eqref{eq:min_error_sdp},  
\begin{multline}
   P^{\rm dual}\!= \max_{Y} \Tr(Y),\\
    \text{subject to}\;
    \frac{\rho_g}{|{\cal G}|} - Y \geq 0, \; \mbox{for all}\; g \in \mathcal{G}.
    \label{eq:min_error_dual_sdp}
\end{multline}  
To maximize \(\Tr(Y)\), we propose a natural choice for the dual operator: 
\( Y = (1/|{\cal G}|) \sum_{g\in{\cal G}} U_g \ket{\omega} \braket{\omega}{\psi} \bra{\psi} U^\dagger_g \),  
ensuring \mbox{\(\Tr(Y) = P^{\rm primal}\)}. With this choice, the (Holevo-like) constraint in Eq.~(\ref{eq:min_error_dual_sdp}) simplifies to \(\ketbra{\psi}{\psi} - |\mathcal{G}|Y \geq 0\), invoking group covariance. This inequality holds since the left-hand side can be written as a convex combination of positive semidefinite operators~\cite{supplemental}.
 This confirms that our choice provides a feasible solution to the dual problem, proving \( P^{\rm dual} = P^{\rm primal} \) and, in turn, validating~Eq.~(\ref{the Pe}).

Next, we address unambiguous QSE, also when the condition \(\sqrt{\lambda_1} > \sum_{a>1} \sqrt{\lambda_a}\) holds.  
With the ansatz~(\ref{eq:non_perf_ansatz}) and choice~(\ref{gamma unamb}), the inconclusive POVM element reads \mbox{$\Pi_Q = \openone-\sum_{g\in\mathcal{G}}U_g|\omega\rangle\langle\omega| U_g^\dag$}.
This specific choice of the seed state $|\omega\rangle$ ensures that \mbox{$\Tr(\Pi_g\rho_g)=|\braket{\omega}{\psi}|^2=0$}, and  \mbox{$\Pi_Q = (1-\gamma^2)\ketbra{v^1_1}{v^1_1}\otimes\ketbra{u^1_1}{u^1_1}\geq 0$},
since we can alternatively write \mbox{$\gamma=(1/\sqrt{\lambda_1})\sum_{a\not=1}\sqrt{\lambda_a}<1$}. 
Thus our ansatz is a feasible solution of the (primal) SDP in Eq.~(\ref{eq:zero_error_sdp}), with the probability of an inconclusive outcome given by
\mbox{$Q^{\mathrm{primal}} =(\alpha^1_1/|{\cal G}|)(1-\gamma^2)$}, which simplifies to our main result in Eq.~(\ref{eq:Pe-unambiguous}).  

To show the optimality of this ansatz, we consider the dual formulation of Eq.~(\ref{eq:zero_error_sdp}),  
\begin{multline}\label{eq:dual_sdp_UA}  
        \kern-.2em Q^{\rm dual}\!:= \max_{X} \;1-\Tr(X), \;\mbox{subject to}\\  
        \sum_{g\in\mathcal{G}}U_gXU_g^\dag + \nu\ketbra{\psi}{\psi}-\Omega \geq 0,\;  
        \nu\in\mathbb{R} ,  
\end{multline}  
which has been particularized for the problem at hand and simplified by exploiting its symmetries.
This maximum is not strictly achievable, as the optimal solution lies on the boundary of the feasibility set, which is not contained within the feasibility set itself~\cite{boyd_2004,watrous_2018}. As a result, unlike in the minimum-error case, there is no natural choice for the dual operator~$X$.  

Nevertheless, strong duality holds, and it is possible to construct $X$ such that the dual problem yields an inconclusive probability satisfying \mbox{$Q^{\mathrm{dual}} = Q^{\mathrm{primal}}-\varepsilon$}, where $\varepsilon$ is an arbitrary positive real number~\cite{supplemental}. This construction ensures that for any such~$\varepsilon$, there exists a corresponding $\nu$ in Eq.~(\ref{eq:dual_sdp_UA}) that makes the solution feasible. Hence, taking the limit $\varepsilon\rightarrow0^+$, we obtain \mbox{$Q^{\mathrm{dual}} = Q^{\mathrm{primal}}=Q$}, proving the optimality of the ansatz~(\ref{eq:non_perf_ansatz}) with the specified~$\gamma$.  
\end{proof}

\noindent\textit{Conclusions}.  
We have obtained the optimal minimum-error and unambiguous quantum state exclusion protocols for any (finite-)group-generated set of states, offering a constructive proof that provides the explicit form of the optimal POVM in both cases. Several previously known results emerge as particular instances of our findings.  
A key aspect of our approach is the systematic use of the Gram matrix, which encodes all the information required for discrimination and exclusion, regardless of whether the set is symmetric or not.  
Our expressions for the failure probabilities, Eqs.~(\ref{the Pe}) and~(\ref{eq:Pe-unambiguous}), depend solely on the Gram matrix eigenvalues, making them applicable even to non-symmetric sets.  
Numerical analysis further suggests that these expressions provide upper bounds on the corresponding probabilities for the optimal QSE protocols, aligning with the heuristic expectation that discrimination and exclusion become more challenging as the states share more symmetries.  Our theoretical treatment can also be extended to group-generated ensembles of mixed states, where perfect exclusion is generally not achievable. In such cases, our approach offers a path to estimate the minimum error probability instead. We are currently exploring these ideas further.

\textit{Acknowledgments}. We are grateful to Prof. Sandu Popescu for valuable discussions and for suggesting the inclusion of projective representations in our study.
This work has been financially supported by MCIN with funding from European Union NextGenerationEU \mbox{(PRTR-C17.I1)} and by Generalitat de Catalunya. We also acknowledge support from the Ministry of Economic Affairs and Digital Transformation of the Spanish Government through the QUANTUM ENIA project: Quantum Spain, by the European Union through the Recovery, Transformation and Resilience Plan - NextGenerationEU within the framework of the “Digital Spain 2026 Agenda”, and by grant PID2022-141283NB-I00 funded by \mbox{MICIU/AEI/10.13039/501100011033}. R.M.T.  acknowledges financial support from MCIN mobility grant PRX23/00600, ERC grant FLQuant, ID: 101021085,
and the kind hospitality of the H.H. Wills Physics Laboratory of the University of Bristol. A.D. also acknowledges support from Ministerio de Ciencia e Innovación of the Spanish Government FPU23/02763.

\textit{Data availability}. No data were created or analyzed in this study.

\bibliography{bibliography} 

\cleardoublepage 

\setcounter{section}{0} 
\setcounter{equation}{0} 
\setcounter{figure}{0} 
\setcounter{table}{0} 
\setcounter{page}{1}

\renewcommand{\theequation}{S\arabic{equation}}

\title{SUPPLEMENTAL MATERIAL: \\ Quantum state exclusion for group-generated ensembles of pure states}
\author{A. Diebra$^{1}$, S. Llorens$^{1}$,  E. Bagan$^{1}$, G. Sent\'{i}s$^{1,2}$, and R. Muñoz-Tapia$^{1,3}$}

\affiliation{$^{1}$F\'{i}sica Te\`{o}rica: Informaci\'{o} i Fen\`{o}mens Qu\`antics, Universitat Aut\`{o}noma de Barcelona, 08193 Bellaterra (Barcelona), Spain
}
\affiliation{$^{2}$Ideaded, Carrer de la Tecnologia, 35, 08840 Viladecans, Barcelona, Spain
}
\affiliation{$^{3}$H. H. Wills Physics Laboratory, University of Bristol, Tyndall Avenue, Bristol, BS8 1TL, United Kingdom
}
\begin{abstract}
These supplementary notes provide a proof of the informational completeness of the Gram matrix in discrimination and exclusion problems, along with a brief summary of finite-group representation theory, including both linear and projective representations, focused on applications relevant to the main text. Additionally, they offer further details on the proofs of the Lemma presented therein.
\end{abstract}

\setcounter{secnumdepth}{2}

\maketitle
\onecolumngrid
All equations in this supplementary note are numbered with the prefix `S'. Equations referenced without this prefix correspond to those in the main text.

\section{Gram Matrix as a Complete Descriptor of Pure State Ensembles}\label{sec 1}  
In this section, we provide a formal proof that, in the context of state discrimination and exclusion, the Gram matrix encapsulates all the information needed to compute averaged cost functions and figures of merit, as well as to determine the corresponding optimal measurements.


More precisely, we consider the task of discriminating or excluding pure states drawn from a given ensemble \( \mathcal{E}:=\{\eta_k, \ket{\psi_k}\}_{k=1}^N \), where the states have prior probabilities \( \eta_k \) and may or may not be linearly independent.
We assume that the states belong to a \( d \)-dimensional Hilbert space, i.e., \( \ket{\psi_k} \in \mathbb{C}^d \). 
The Gram matrix \( G \) associated with~\( \mathcal{E} \) is defined as  
\begin{equation}
\label{gram general}
G_{k,l} = \sqrt{\eta_k \eta_l} \braket{\psi_k}{\psi_l}.
\end{equation}
It can be viewed as an operator on \( \mathbb{C}^N \), given by  
\begin{equation}
\label{gram operator general}
G = \sum_{k,l=1}^N \sqrt{\eta_k \eta_l} \braket{\psi_k}{\psi_l} \ketbra{k}{l},
\end{equation}
where \( \{\ket{k}\}_{k=1}^N \) is an arbitrary orthonormal basis, which may be chosen to suit the specific problem at hand.

The cost functions relevant to discrimination and exclusion tasks generically take the form  
\begin{equation}\label{eq_app:figure_of_merit}
    \mathcal{F} = \sum_{l,k=1}^N \eta_k f_{k,l} \bra{\psi_k} \Pi_l \ket{\psi_k},
\end{equation}  
where \( \{\Pi_k\}_{k=1}^N \) defines the Positive Operator-Valued Measure (POVM) describing the quantum measurement performed on \( \ket{\psi_k} \) to carry out the task. The joint probability of preparing the state \( \ket{\psi_k} \) from \( \mathcal{E} \) and obtaining measurement outcome \( l \) is given by \( \eta_k \bra{\psi_k} \Pi_l \ket{\psi_k} \). Thus, the choice of coefficients \( f_{k,l} \) determines the (averaged) cost function; for instance, in the exclusion problem, the average error probability is obtained by setting \( f_{k,l} = \delta_{k,l} \).
The operators~\( \Pi_k \) are positive semidefinite and satisfy  
\begin{equation}\label{eq_app:POVM_condition}
    \openone_d - \sum_{k=1}^N \Pi_k \geq 0,
\end{equation}  
where \( \openone_d \), the $d$-dimensional identity operator (or matrix), may be replaced by the projector onto the span of \( \mathcal{E} \), denoted \( \openone_{\mathrm{span}(\mathcal{E})} \) (obviously, \( \openone_d - \openone_{\mathrm{span}(\mathcal{E})} \geq 0 \)). 
Additional constraints of the form  
\begin{equation}\label{constraint form}
\sum_{k,l=1}^{N} c^a_{l,k} \langle \psi_k | \Pi_l | \psi_k \rangle = 0, \qquad a = 1,2,\dots
\end{equation}
may apply depending on the problem, where \( c^a_{l,k} \) are fixed coefficients. For instance, in unambiguous exclusion, one requires  
\( \Tr(\Pi_k \rho_k) = \langle \psi_k | \Pi_k | \psi_k \rangle = 0 \) for all \( k \); see Eq.~(\ref{eq:zero_error_sdp}).

The following theorem formalizes the claim that the Gram matrix fully describes a pure state ensemble for computing averages in state discrimination and exclusion:

\begin{theorem}\label{theorem_all_u_need_is_gram}
    For any POVM $\{\Pi_k\}_{k=1}^N$ on $\mathbb{C}^d$, there exists a set of positive operators $\{\Xi_k\}_{k=1}^N$ acting on $\mathbb{C}^N$ satisfying
    \begin{equation}\label{eq_app:completeness_F}
        G - \sum_{k=1}^N \Xi_k \geq 0,
    \end{equation}
    such that 
    \begin{equation}\label{eq_app:expval_equivalence_F}
        \eta_k\bra{\psi_k}\Pi_l\ket{\psi_k} = \bra{k}\Xi_l\ket{k}.
    \end{equation}
    Conversely, given a set of positive operators $\{\Xi_k\}_{k=1}^N$ on $\mathbb{C}^N$ satisfying Eq. (\ref{eq_app:completeness_F}), there exists a POVM on $\mathbb{C}^d$ for which Eq. (\ref{eq_app:expval_equivalence_F}) holds.
\end{theorem}
%
%

\begin{adjustwidth}{2em}{0em}
{\em Proof.} Define the operator
\begin{equation}\label{eq_app:X_operator}
    X = \sum_{k=1}^N \sqrt{\eta_k} \ketbra{\psi_k}{k},
\end{equation}
which satisfies \( G = X^\dagger X \). Given a POVM \( \{\Pi_k\}_{k=1}^N \), define \( \Xi_k = X^\dagger \Pi_k X \geq 0 \). Then, the POVM condition in Eq.~(\ref{eq_app:POVM_condition}) becomes
\begin{equation}
    X^\dagger X - \sum_{k=1}^N X^\dagger \Pi_k X \geq 0,
\end{equation}
proving Eq.~(\ref{eq_app:completeness_F}). On the other hand, since \( X \ket{k} = \sqrt{\eta_k} \ket{\psi_k} \), it follows that
\begin{equation}
    \eta_k \bra{\psi_k} \Pi_l \ket{\psi_k} = \bra{k} X^\dagger \Pi_l X \ket{k} = \bra{k} \Xi_l \ket{k}.
\end{equation}
This completes the proof of the direct statement.

To prove the converse, let \( X^{+} \) denote the Moore-Penrose pseudoinverse of \( X \). Given positive semidefinite operators~\( \{\Xi_k\}_{k=1}^N \) that satisfy Eq.~(\ref{eq_app:completeness_F}), define
\begin{equation}
    \Pi_k := (X^{+})^\dag \Xi_k X^{+} \geq 0.
\end{equation}
Multiplying Eq.~(\ref{eq_app:completeness_F}) by \( (X^{+})^\dag \) from the left and by \( X^{+} \) from the right yields
\begin{equation}\label{eq_app:almost_POVM_condition}
    (X^{+})^\dag X^\dag XX^{+} - \sum_{k=1}^N (X^{+})^\dag \Xi_k X^{+} \geq 0.
\end{equation}
Noticing that \( XX^{+} = \openone_{\mathrm{range}(X)} \) is the projector onto \( \mathrm{range}(X) = \mathrm{span}(\mathcal{E}) \), one obtains that the first term of this expression equals \( \openone_{\Span(\mathcal{E})} \leq \openone_d \), recovering the POVM condition~(\ref{eq_app:POVM_condition}). Additionally
\begin{equation}
        \eta_k \bra{\psi_k} \Pi_l \ket{\psi_k} = \bra{k} X^\dag \Pi_l X \ket{k} = \bra{k} \left[ (X^{+} X)^\dag \Xi_l X^{+} X \right] \ket{k} = \bra{k} \Xi_l \ket{k},
\end{equation}
where, in the final equality, we used \( X^{+}X = \openone_{\Range(X^\dag)} \), \( \Range(X^\dag) = \Range(G) \), and the fact that \( \Xi_k \openone_{\Range(G)} = \Xi_k \), which follows from condition~(\ref{eq_app:completeness_F}). This completes the proof~$\blacksquare$
\end{adjustwidth}

\noindent Using Theorem~\ref{theorem_all_u_need_is_gram}, we can rewrite the averaged cost function in Eq.~(\ref{eq_app:figure_of_merit}) as  
\begin{equation}\label{eq_app:figure_of_merit II}
    \mathcal{F} = \sum_{l,k=1}^N  f_{k,l} \langle k|\Xi_l|k\rangle,
\end{equation}  
with the POVM condition~(\ref{eq_app:POVM_condition}) reexpressed in terms of the positive semidefinite operators \( \{\Xi_k\}_{k=1}^N \) as in Eq.~(\ref{eq_app:completeness_F}).
%
%
Similarly, if additional constraints such as those in Eq.~(\ref{constraint form}) are imposed, they can be written as  
\begin{equation}\label{constraint form II}
\sum_{k,l=1}^N \tilde{c}^a_{l,k} \langle k|\Xi_l|k\rangle = 0, \qquad a=1,2,\dots
\end{equation}
for some new coefficients $ \tilde{c}^a_{l,k}$.
Thus, the various elements needed to formulate the discrimination or exclusion problem, originally expressed in Eqs.~(\ref{eq_app:figure_of_merit})--(\ref{constraint form}), can be equivalently reformulated as Eqs.~(\ref{eq_app:figure_of_merit II})--(\ref{constraint form II}), where the full description of the ensemble \( \mathcal{E} \) is encoded in the Gram matrix \( G \).

Although not widely used in the context of discrimination and exclusion, it also follows from Theorem~\ref{theorem_all_u_need_is_gram} that the Gram matrix remains a sufficient descriptor of ensembles even for non-linear cost functions, such as mutual information, and for non-linear constraints, as long as both depend solely on the joint probabilities \( \eta_k \bra{\psi_k} \Pi_l \ket{\psi_k} \).

Another direct consequence of Theorem \ref{theorem_all_u_need_is_gram} is that any two ensembles \( \mathcal{E} = \{\eta_k, \ket{\psi_k}\}_{k=1}^N \) and \( \tilde{\mathcal{E}} = \{\tilde\eta_k, \ket{\tilde\psi_k}\}_{k=1}^N \), that share the same Gram matrix, will exhibit identical discrimination and exclusion properties. Moreover, for any given POVM \( \{\Pi_k\}_{k=1}^N \), there exists an alternative one \( \{\tilde{\Pi}_k\}_{k=1}^N \) such that 
\begin{equation}
    \eta_k \bra{\psi_k} \Pi_l \ket{\psi_k} = \tilde{\eta}_k \bra{\tilde\psi_k} \tilde{\Pi}_l \ket{\tilde\psi_k} \quad \text{for all} \quad k, l.
\end{equation}
Such a POVM can be constructed as 
\begin{equation}
    \tilde{\Pi}_k := (X \tilde{X}^+)^\dag \Pi_k X \tilde{X}^+,
\end{equation}
where \( X \) is defined in Eq.~(\ref{eq_app:X_operator}) and \( \tilde{X} \) is defined similarly as 
\begin{equation}
    \tilde{X} = \sum_{k=1}^N \sqrt{\tilde{\eta}_k} \ketbra{\tilde{\psi}_k}{k}.
\end{equation}

Lastly, noticing that the ensemble operator \( \Omega := \sum_k \eta_k \ketbra{\psi_k}{\psi_k} \) can be written as \( \Omega = XX^\dag \), one concludes that both the Gram matrix \( G \) and the ensemble operator \( \Omega \) share the same non-zero eigenvalues —a result that has been extensively used in the main text.

\section{Representation Theory of Finite Groups: A Concise Summary}

This section provides an overview of key concepts in representation theory extensively used throughout the letter. The results summarized here can be found in standard textbooks \cite{hamermesh1989group,steinberg2011representation} and also in review articles such as \cite{christandl2006structurebipartitequantumstates,CHENG2015230}.

\subsection{Linear representations}\label{lin repr 1}

A \textit{representation} of a group \( \mathcal{G} \) is a homomorphism from \( \mathcal{G} \) to the group of invertible linear transformations (matrices) acting on a vector space \( V \). Representations provide a powerful way to study the structure and properties of abstract groups through linear algebra. In addition, unitary representations are central in quantum physics, as they describe transformations of physical systems.

Formally, if \( GL(V) \) denotes the general linear group on \( V \), a representation is a map \mbox{\( D: \mathcal{G} \to GL(V) \)} satisfying  
\begin{equation}
   D(g) D(h)= D(gh) , \quad D(e) = \openone,
\end{equation}
for all \( g, h \in \mathcal{G} \), where \( e \in \mathcal{G} \) is the identity element and \( \openone \) is the identity matrix in \( GL(V) \).

In these notes, we focus exclusively on \textit{unitary representations}, where each group element is mapped to a unitary matrix. That is, we consider maps \mbox{\( U: \mathcal{G} \to \mathcal{U}(V) \)}, where \( \mathcal{U}(V) \subset GL(V) \) is the group of unitary transformations on~\( V \), which is typically a Hilbert space \( \mathcal{H} \) in quantum mechanical applications. In this case, it is customary to write~\( U \) instead of \( D \) and use the subscript notation \( U_g := U(g) \in \mathcal{U}(V) \).


A central concept in representation theory is irreducibility. A representation \( U \) is said to be \textit{irreducible} if the only subspaces \( W \subseteq V \) that remain invariant under the action of all \( U_g \) are the trivial subspaces, namely the zero subspace \mbox{\( W = \{0\} \)} and the entire space \mbox{\( W = V \)}. Irreducible representations, often referred to as irreps, are fundamental because any representation \( U \) of a finite group \( \mathcal{G} \) is either irreducible or can be decomposed into a direct sum of irreps, in which case it is called {\em reducible}. Specifically, we have
\begin{equation}
\label{irrep decomposition}
    U \cong \bigoplus_{\mu} m_\mu U^\mu,
\end{equation}
where \( \mu \) labels the distinct irreps \( U^\mu \) of \( \mathcal{G} \) and \( m_\mu \) denotes their multiplicity.
In matrix terms, this means that by choosing an appropriate basis, each group element \( U_g \) can be written as
\begin{equation}
\label{irrep decomposition 2}
    U_g = \bigoplus_\mu U_g^\mu \otimes \openone_{m_\mu},
\end{equation}
where \( \openone_{m_\mu} \) accounts for the multiplicity of the irreps.
The irreps of a group are fundamentally determined by its structure. Each finite group has a finite set of irreps, with each irrep corresponding to a conjugacy class of the group. 

A fundamental result in the classification of irreducible representations is {\em Schur's lemma}. In its matrix form, it states that given two irreps, \( U_g^\mu \) and \( U_g^\nu \), if a matrix \( T \) satisfies \( T U_g^\mu = U_g^\nu T \) for all \( g \in \mathcal{G} \), then \( T \) is either invertible —implying that the two representations are isomorphic— or zero. Moreover, in the special case where both irreps coincide, \( \mu = \nu \), i.e., when \( T \) commutes with \( U_g^\mu \) for all \( g \in \mathcal{G} \), \( T \) must be a scalar multiple of the identity in the corresponding irreducible subspace. In other words, \( T = \lambda \openone_{d_\mu} \) for some \( \lambda \in \mathbb{C} \), where \( d_\mu \) is the dimension of the irrep.

A direct consequence of Schur's lemma is that the matrix elements of two irreps satisfy the following orthogonality relation, known as the {\em Great Orthogonality Theorem},
\begin{equation}\label{eq_app:great_ortho_relation}
\sum_{g\in\mathcal{G}}\left(U_g^\mu\right)_{n,m}\left(U_g^\nu\right)_{n',m'}^* = \delta_{\mu,\nu}\delta_{n,n'}\delta_{m,m'}\frac{|\mathcal{G}|}{d_\mu},    
\end{equation}
where ${}^*$ denotes complex conjugation, the sum extends over all group elements \( g \in \mathcal{G} \), and \( |\mathcal{G}| \) is the order of the group.

A well-known example of a reducible representation is the {\em regular representation}, which plays an importantl role in the proof of the lemma in the main text. This representation arises from the natural action of \( \mathcal{G} \) on itself. Specifically, each element \( g \in \mathcal{G} \) is associated with an orthonormal basis \( \{ \ket{g} \}_{g \in \mathcal{G}} \) of a \( |\mathcal{G}| \)-dimensional Hilbert space, commonly denoted as \( \mathbb{C}[\mathcal{G}] \) and referred to as the {\em group algebra}. The {\em left-(right-)regular representation} is then defined by left (right) translations:  
\begin{equation}
    L_g \ket{c} = \ket{gc},\quad R_g \ket{c} = \ket{cg^{-1}}.
\end{equation}
%
%
The matrices of \( L_g \) and \( R_g \) are \mbox{\((0,1)\)-matrices} (i.e., matrices with binary entries). Their matrix elements are given by 
\begin{equation}
\label{matrices L&R}
(L_g)_{r,c} = \delta_{g,rc^{-1}},\qquad (R_g)_{r,c} = \delta_{g,r^{-1}c}\,, 
\end{equation}
where \( \delta_{g,h} \) is the Kronecker delta over the elements of \( \mathcal{G} \). 
 
These matrices can be directly constructed from the group's multiplication table. Specifically, consider a Cayley~table \( \mathscr{L} \) where the column headings list the inverses of the group elements instead of the elements themselves. In this table, each entry is given by \( \mathscr{L}_{r,c} = rc^{-1} \). Similarly, define \( {\mathscr{R}} \) as a Cayley table where the row headings list the inverses, so that \( \mathscr{R}_{r,c} = r^{-1}c \).  
Comparing with Eq.~(\ref{matrices L&R}), we see that the matrix \( L_g \) (resp. \( R_g \)) is obtained by replacing every occurrence of \( g \) in \( \mathscr{L} \) (resp. \({\mathscr{R}} \)) with 1, while all other entries are replaced by~0.

Both regular representations are unitary, and their traces satisfy \( \Tr(L_g) = \Tr(R_g) = |\mathcal{G}| \delta_{g,e} \), meaning that they are traceless matrices except for the identity element \( e \). As a result, it can be shown from character theory that the decomposition of the regular representations contains all the distinct irreps of the group, each appearing with multiplicity equal to its dimension:
\begin{equation}
\label{regular_decomp}
    L \cong \bigoplus_{\mu} d_\mu U^\mu, \quad R  \cong \bigoplus_{\mu} d_\mu U^\mu.
\end{equation}
Taking traces, we obtain the well-known result
\begin{equation}
\label{order dim}
    |\mathcal{G}| = \sum_{\mu} d_\mu^2,
\end{equation}
which relates the order of the group to the dimensions of its irreps. An immediate consequence is that all irreps of a finite abelian group are one-dimensional: since all elements of an abelian group commute, the number of conjugacy classes —and hence the number of distinct irreps— equals the order of the group, which, by Eq.~(\ref{order dim}), forces \( d_\mu = 1 \) for all \( \mu \).

\subsection{Projective representations}\label{proj repr}
Projective representations generalize linear representations by allowing multiplicative prefactors in the composition rule of \( D \). Specifically, a {\em projective representation} of a group \( \mathcal{G} \) on a vector space \( V \) is a map \( D: \mathcal{G} \to GL(V) \) satisfying  
\begin{equation}
   D(g) D(h) = \omega(g,h)  D(gh), \quad D(e) = \openone,
\end{equation}  
for all \( g,h \in \mathcal{G} \), where \( \omega(g,h) \in \mathbb{C}^\times \) are nonzero complex functions of the group elements and \(\mathbb{C}^\times\) is the multiplicative group of complex numbers.

Such prefactors are not arbitrary; the map \( \omega: \mathcal{G} \times \mathcal{G} \to \mathbb{C}^\times \) must satisfy  
\begin{equation}\label{multiplier associativity cond}  
    \omega(g,h)\omega(gh,f) = \omega(g,hf)\omega(h,f),\quad   \omega(g,e)=\omega(e,g)=1,\quad \mbox{for all $g,h,f\in\mathcal{G}$,}  
\end{equation}  
where the first relation ensures associativity. 
The map \( \omega \) is commonly referred to as a \emph{multiplier} (or 2-cocycle in topology terminology). A projective representation is then completely determined by the tuple \( (D, \omega) \).  

The introduction of multipliers broadens the concept of equivalence between representations. We say that {\em two multipliers \( \omega \) and \( \omega' \) are equivalent} if there exists a map \( \mu: \mathcal{G} \to \mathbb{C}^\times \) with \( \mu(e) = 1 \) such that  
\begin{equation}\label{multiplier equivalence}  
    \omega(g,h) = \frac{\mu(g)\mu(h)}{\mu(gh)} \omega'(g,h), \qquad \mbox{for all $g,h \in \mathcal{G}$.}  
\end{equation}  
It is straightforward to verify that Eq.~(\ref{multiplier equivalence}) defines an equivalence relation.  
{\em Two projective representations} \( (D, \omega) \) {\em and} \( (D', \omega') \) {\em are said to be equivalent} if \( D \) and \( D' \) are isomorphic, meaning that there exists an invertible matrix \( T \) such that \( D'(g)\! =\! T D(g) T^{-1} \) for all \( g \in \mathcal{G} \), and their corresponding multipliers are equivalent. Then, $D'(g)\!=\![1/\mu(g)]TD(g)T^{-1}$.
%

It is sometimes possible to reduce a projective representation \( (D, \omega) \) to a linear representation. This is always the case if there exists a map \( \mu: \mathcal{G} \to \mathbb{C}^\times \) with \( \mu(e) = 1 \) such that \( \omega \) takes the form  
\begin{equation}  
    \omega(g,h) = \frac{\mu(g)\mu(h)}{\mu(gh)}, \qquad \mbox{for all $g,h \in \mathcal{G}$.}  
\end{equation}  
When \( \omega \) has this form, we say that \( \omega \) is a \textit{2-coboundary}. From Eq.~(\ref{multiplier equivalence}), we immediately see that such an \( \omega \) is equivalent to the trivial multiplier \( \omega'(g,h) = 1 \) for all \( g,h \in \mathcal{G} \). A projective representation \( (D, \omega) \) where \( \omega \) is a 2-coboundary is referred to as a \emph{trivial (projective) representation}, as it can be reduced to a linear representation \( D' \) defined by~\( D'(g) = [1/\mu(g)] D(g) \):  
\begin{equation}  
    D'(g) D'(h) = \frac{1}{\mu(g)\mu(h)} D(g) D(h) = \frac{1}{\mu(gh)} D(gh) = D'(gh).  
\end{equation}  
Consequently, all one-dimensional projective representations \mbox{\( \chi: \mathcal{G} \to \mathbb{C} \), for which \( \chi(g)\chi(h) = \omega(g,h) \chi(gh) \)}, are trivial, as the corresponding multiplier \( \omega \) is clearly a 2-coboundary, with \( \mu(g) = \chi(g) \).

As with linear representations, we focus exclusively on \textit{unitary projective representations} \( U: \mathcal{G} \to \mathcal{U}(V) \), where the corresponding multipliers are pure phases (unitary), i.e., 
$\omega(g,h)=e^{i\theta(g,h)}$, $\theta(g,h)\in\mathbb{R}$ for all \( g,h \in \mathcal{G} \). Given any projective representation $D$, one can always find an equivalent unitary projective representation.  

In quantum mechanics, where we are primarily concerned with the action of unitary representations on quantum states~\mbox{\( |\psi\rangle \in \mathcal{H} \)}, global phases are irrelevant since physical states correspond to rays in Hilbert space. This freedom implies that \( U_g \) and \( U'_g = \mu(g) U_g \) are physically equivalent for any pure phase \( \mu(g) \). Consequently, equivalent projective unitary representations, as defined in Eq.~(\ref{multiplier equivalence}), induce the same physical transformations on a quantum system.

Most properties of linear representations naturally extend to projective representations when properly accounting for the multiplier \( \omega \). In particular, a reducible projective representation \( (U,\omega) \) decomposes as a direct sum of irreducible projective representations, as in Eqs.~(\ref{irrep decomposition}) and (\ref{irrep decomposition 2}), satisfying \( U_g^\mu U_h^\mu = \omega(g,h) U_{gh}^\mu \) with the same multiplier \( \omega \) for all \( \mu \).  
Furthermore, Schur's lemma applies to projective representations without modification, while the Great Orthogonality Theorem now reads 
\begin{equation}  
\label{GOT projective}
    \sum_{g\in\mathcal{G}}\omega^*(g,g^{-1})\left(U_g^\mu\right)_{n,m}\big(U_{g^{-1}}^\nu\big)_{m',n'} = \delta_{\mu,\nu}\delta_{n,n'}\delta_{m,m'}\frac{|\mathcal{G}|}{d_\mu}.    
\end{equation} 
Note that while Eq. (\ref{eq_app:great_ortho_relation}) involves \( U_g^\dag \), for projective representations, \( U_{g^{-1}} \) appears instead, and in general, \( U_{g^{-1}} \neq U_g^\dag \).  

Having provided a general overview of projective representations of groups, we now turn to considerations specific to the objectives of the main text. In particular, we aim to derive a canonical form for the Gram matrix (Sec.~\ref{projective Gram} below) and show that our results, especially Eqs.~(\ref{the Pe}) and~(\ref{eq:Pe-unambiguous}), also hold in the case of projective representations.

For general unitary projective representations, it has been noted that \( U_{g^{-1}} \neq U_g^\dag \), meaning that the inverse of a matrix does not necessarily belong to the representation. However, the two matrices are related by
\begin{equation}
    U_g^\dag = \omega^*(g,g^{-1}) U_{g^{-1}} = \omega^*(g^{-1},g) U_{g^{-1}},
\end{equation}
which follows from the relation \( U_g U_{g^{-1}} = \omega(g,g^{-1}) U_{gg^{-1}} = \omega(g,g^{-1}) \openone \) and the analogous expression for \( U_{g^{-1}} U_g \).
Defining the map \mbox{$\mu: \mathcal{G} \to \mathbb{C}^\times$} by 
\begin{equation}\label{mu=1/sqrt w}
\mu(g) = {1\over\sqrt{\omega(g,g^{-1})}}
\end{equation}
(choosing either branch of the square root), we see that the equivalent representation \mbox{$U'_g = \mu(g) U_g$} satisfies \mbox{$U'_g U'_{g^{-1}} = \openone$}, i.e., $(U'_g)^\dag = U'_{g^{-1}}$.  
Thus, when the multipliers are unitary (pure phases), we may assume without loss of generality that 
\begin{equation}\label{assump gg-1}
\omega(g, g^{-1}) = \omega(g^{-1},g) =1,\qquad \mbox{for all \( g \in \mathcal{G}\). }
\end{equation}
Under this assumption, the Great Orthogonality Theorem in Eq.~(\ref{GOT projective}) takes the same form as its counterpart for linear representations in Eq.~(\ref{eq_app:great_ortho_relation}).

The concept of the regular representation extends to projective representations. In analogy with linear representations, the {\em projective left- and right-regular representations} with multiplier \( \omega \) are defined by left and right translations over the group algebra:
\begin{equation}\label{projective regular}
    L_g\ket{h} = \omega(g,h)\ket{gh}, \quad
    R_g\ket{h} = \omega(g,h^{-1})\ket{hg^{-1}}.
\end{equation}
For unitary \( \omega \), both representations are unitary, though they are no longer \mbox{\((0,1)\)-matrices}; instead, their entries are pure phases:
\begin{equation}
\label{matrices L&R proj}
(L_g)_{r,c} = \omega(rc^{-1},c) \delta_{g,rc^{-1}} = [\omega(r,c^{-1})]^* \delta_{g,rc^{-1}}, \quad 
(R_g)_{r,c} = \omega(r^{-1}c,c^{-1}) \delta_{g,r^{-1}c} = [\omega(r^{-1},c)]^* \delta_{g,r^{-1}c},
\end{equation}
where we have used Eq.~(\ref{multiplier associativity cond}) and assumption~(\ref{assump gg-1}).
The traces of these matrices satisfy \( \Tr(L_g) = \Tr(R_g) = |\mathcal{G}|\delta_{g,e} \), implying that the projective regular representations obeys Eqs.~(\ref{regular_decomp}) and~(\ref{order dim}).

As in the case of linear representations, the matrices of the two regular representations can be systematically derived from ``Cayley-like tables" that also incorporate the multipliers: 
\begin{equation}\label{formal Cayley}
\mathscr{L}_{r,c} = \omega(r, c^{-1}) r c^{-1}, \quad \mathscr{R}_{r,c} = \omega(r^{-1}, c) r^{-1} c,
\end{equation}
where the right-hand sides are formal expressions used for bookkeeping purposes. Comparing with Eq.~(\ref{matrices L&R proj}), the matrix of \( L^*_g \) ---the complex conjugate of the left-regular representation, as defined in Eq.~(\ref{projective regular})--- can be obtained from the table \(\mathscr{L}\) by replacing all occurrences of \( g \) with 1 and setting the remaining entries to 0. To construct \( L_g \), we then take the complex conjugate of the resulting matrix. Similarly, applying the same procedure to the multiplication table \(\mathscr{R}\) yields \( R^*_g \) and \( R_g \).
The sets \( \{L_g^*\}_{g\in\mathcal{G}} \) and \( \{R_g^*\}_{g\in\mathcal{G}} \) themselves form regular representations.

\section{Group-generated Gram matrices}

This section discusses the structure and construction of Gram matrices associated with group-generated pure state ensembles. The key insight is that the symmetry group generating the ensemble fully determines the structure of its Gram matrix. We illustrate our findings with concrete examples.

\subsection{Linear representations}

Consider the ensemble \( \mathcal{E} = \{\ket{\psi_g} = U_g \ket{\psi} \,|\, g \in \mathcal{G}\} \) of equiprobable states generated by a group \( \mathcal{G} \). To simplify the analysis, we omit  the prior probability \( \eta_g = 1/|\mathcal{G}| \) from the definition of the associated Gram matrix, Eq.~(\ref{gram general}), and instead define \( G_{g,h} = \braket{\psi_g}{\psi_h} \). Identifying the orthonormal basis in Eq.~(\ref{gram operator general}) with the basis of the group algebra, we obtain  
\begin{equation}
    G = \sum_{g,h\in\mathcal{G}}\braket{\psi_g}{\psi_h}\ketbra{g}{h} = \sum_{g,h\in\mathcal{G}}\braket{\psi}{\psi_{g^{-1}h}}\ketbra{g}{h} = \sum_{l,h\in\mathcal{G}}\braket{\psi}{\psi_l}\ketbra{hl^{-1}}{h} = \sum_{l,h\in\mathcal{G}}\braket{\psi}{\psi_l}R_l\ketbra{h}{h} = \sum_{l\in\mathcal{G}}\braket{\psi}{\psi_l}R_l.
\end{equation}  
Thus, the Gram matrix \( G \) is contained in the associative algebra spanned by the right-regular representation of the group that generates the ensemble, fully determining its structure.

The above provides a direct and systematic procedure for constructing Gram matrices with a desired symmetry inherited from the group~$\mathcal{G}$, bypassing the explicit construction of the associated ensemble $\mathcal{E}$. The steps are as follows:  
\begin{enumerate}[label={(\alph*)},nosep]
\item\label{gram 1} Construct the group's multiplication table $\mathscr{R}$ (see Sec.~\ref{lin repr 1}) listing the elements of $\mathcal{G}$ as column headings and their inverses as row headings.  
\item\label{gram 2} Define a matrix $\tilde G$ by replacing each distinct entry of the table $\mathscr{R}$ with a complex coefficient $c_g$, assigning $c_e = 1$ for the identity element $e$.  
\item\label{gram 3} Impose semidefiniteness constraints on $\tilde G$ to ensure it represents a valid Gram matrix.  
\end{enumerate}
The resulting matrix $G$ has  the desired group symmetry. This procedure is illustrated in the following example.

\begin{adjustwidth}{2em}{0em}
{\small\rm
\noindent{\em Example.} Consider the smallest non-abelian group: the dihedral group $D_3$. It is the symmetry group of an equilateral triangle with vertices A, B, C. The group consists of six elements: three (clockwise) rotations, $e$, $r$, and $r^2$, by angles~$0$, $2\pi/3$, and $4\pi/3$ radians, respectively ($r^3 = e$), and three reflections, $s_{\mathrm{A}}$, $s_{\mathrm{B}}$, and $s_{\mathrm{C}}$, along the lines passing through vertex~A, B, or~C, respectively, and the midpoint of the opposite side.
Step~\ref{gram 1} yields the following multiplication table:

\begin{center}
    \begin{tblr}{
  vline{2} = {-}{},
  hline{2} = {-}{},
}
  $D_3$ & $e$   & $r$ & $r^2$ & $s_{\mathrm{A}}$ & $s_{\mathrm{B}}$ & $s_{\mathrm{C}}$ \\
  $e$   & $e$   & $r$ & $r^2$ & $s_{\mathrm{A}}$ & $s_{\mathrm{B}}$ & $s_{\mathrm{C}}$ \\
  $r^2$ & $r^2$ & $e$   & $r$ & $s_{\mathrm{C}}$ & $s_{\mathrm{A}}$ & $s_{\mathrm{B}}$ \\
  $r$ & $r$ & $r^2$ & $e$   & $s_{\mathrm{B}}$ & $s_{\mathrm{C}}$ & $s_{\mathrm{A}}$ \\
  $s_{\mathrm{A}}$ & $s_{\mathrm{A}}$ & $s_{\mathrm{C}}$ & $s_{\mathrm{B}}$ & $e$   & $r^2$ & $r$ \\
  $s_{\mathrm{B}}$ & $s_{\mathrm{B}}$ & $s_{\mathrm{A}}$ & $s_{\mathrm{C}}$ & $r$ & $e$   & $r^2$ \\
  $s_{\mathrm{C}}$ & $s_{\mathrm{C}}$ & $s_{\mathrm{B}}$ & $s_{\mathrm{A}}$ & $r^2$ & $r$ & $e$
\end{tblr}
\end{center}
In step~\ref{gram 2}, we substitute $r^k \mapsto c_k \in \mathbb{C}$ and $e \mapsto 1$, along with $s_\alpha \mapsto c_\alpha\in\mathbb{C}$, yielding:
\begin{equation}
    \tilde{G} = 
  \begin{pmatrix}
  1   & c_1 & c_2 & c_{\mathrm{A}} & c_{\mathrm{B}} & c_{\mathrm{C}} \\
  c_2 & 1   & c_1 & c_{\mathrm{C}} & c_{\mathrm{A}} & c_{\mathrm{B}} \\
  c_1 & c_2 & 1   & c_{\mathrm{B}} & c_{\mathrm{C}} & c_{\mathrm{A}} \\
  c_{\mathrm{A}} & c_{\mathrm{C}} & c_{\mathrm{B}} & 1   & c_2 & c_1 \\
  c_{\mathrm{B}} & c_{\mathrm{A}} & c_{\mathrm{C}} & c_1 & 1   & c_2 \\
  c_{\mathrm{C}} & c_{\mathrm{B}} & c_{\mathrm{A}} & c_2 & c_1 & 1 
    \end{pmatrix} .
\end{equation}
Finally, in step~\ref{gram 3}, we enforce the conditions required for $G$ to be a valid Gram matrix: hermiticity imposes $c_2 = c_1^*$ and $c_{\mathrm{A}}, c_{\mathrm{B}}, c_{\mathrm{C}} \in \mathbb{R}$, while positive semidefiniteness further constrains the coefficients. Thus,  

\begin{equation}
   G = 
  \begin{pmatrix}
  1   & c_1 & c_1^* & c_{\mathrm{A}} & c_{\mathrm{B}} & c_{\mathrm{C}} \\
  c_1^* & 1   & c_1 & c_{\mathrm{C}} & c_{\mathrm{A}} & c_{\mathrm{B}} \\
  c_1 & c_1^* & 1   & c_{\mathrm{B}} & c_{\mathrm{C}} & c_{\mathrm{A}} \\
  c_{\mathrm{A}} & c_{\mathrm{C}} & c_{\mathrm{B}} & 1   & c_1^* & c_1 \\
  c_{\mathrm{B}} & c_{\mathrm{A}} & c_{\mathrm{C}} & c_1 & 1   & c_1^* \\
  c_{\mathrm{C}} & c_{\mathrm{B}} & c_{\mathrm{A}} & c_1^* & c_1 & 1 
    \end{pmatrix}; \quad c_1\in\mathbb{C}, \quad c_{\mathrm{A}}, c_{\mathrm{B}}, c_{\mathrm{C}}\in\mathbb{R}, \quad \text{such that} \quad G\geq 0.
\end{equation}
}
\end{adjustwidth}

\noindent Since the Gram matrix $G$ belongs to the associative algebra spanned by the set $\{R_g\}_{g\in\mathcal{G}}$, any function of $G$, including its square root $S = \sqrt{G}$, also lies in the same algebra. Consequently,  

\begin{equation}
    S = \sum_{g\in\mathcal{G}} s_g R_g,
\end{equation}

for some complex coefficients $\{s_g\}_{g\in\mathcal{G}}$.


Associated with any ensemble $\mathcal{E}$ that can be represented by the Gram matrix $G$, one can always construct a new ensemble $\mathcal{E}_S=\{\ket{\phi_g},g\in\mathcal{G}\}$ by regarding each column of $S$ as a state, $\ket{\phi_g} = \sum_{h\in\mathcal{G}}S_{h,g}\ket{h}\in\mathbb{C}^{|\mathcal{G}|}$. By construction, the ensemble $\mathcal{E}_S$ is represented by the same Gram matrix, i.e., $G_{g,h}=\braket{\psi_g}{\psi_h}=\braket{\phi_g}{\phi_h}$. Consequently, $\mathcal{E}$ and $\mathcal{E}_S$ share the same (anti-)distinguishability properties, as discussed in Sec.~\ref{sec 1}. The ensemble $\mathcal{E}_S$ is also group-generated by~$\mathcal{G}$ and transforms under the left regular representation:
\begin{equation}
    L_g\ket{\phi_h} = \sum_{l\in\mathcal{G}}S_{l,h}L_g\ket{l} = 
    \sum_{l\in\mathcal{G}}S_{l,h}\ket{gl} =
    \sum_{l\in\mathcal{G}}S_{g^{-1}l,h}\ket{l} = \ket{\phi_{gh}},
\end{equation}
where the last equality follows from \( S_{g^{-1}l,h} = S_{l,gh} \), since \( S \) belongs to the algebra of the right-regular representation:
\begin{equation}
\label{S belongs to R}
S_{g^{-1}l,h}=\sum_{f\in\mathcal{G}} s_f (R_f)_{g^{-1}l,h}
=\sum_{f\in\mathcal{G}} s_f \delta_{f,l^{-1}gh}
=\sum_{f\in\mathcal{G}} s_f (R_f)_{l,gh}
=S_{l,gh}.
\end{equation}  

In particular, the above proves that any matrix constructed by the above procedure is the Gram matrix of at least one group-generated ensemble.

\subsection{Projective representations}\label{projective Gram}

When the ensemble \mbox{\( \mathcal{E} = \{\ket{\psi_g} = U_g \ket{\psi} \,|\, g \in \mathcal{G}\} \)} is generated by a projective representation of a group \( \mathcal{G} \) with multiplier \( \omega \), the associated Gram matrix can be brought into a canonical form that also exhibits a well-defined structure determined by the representation.

We first write the Gram matrix in the basis of the group algebra, obtaining
\begin{multline}
    G = \sum_{g,h\in\mathcal{G}}\braket{\psi_g}{\psi_h}\ketbra{g}{h}  
    = \sum_{g,h\in\mathcal{G}}\omega(g^{-1},h)\braket{\psi}{\psi_{g^{-1}h}}\ketbra{g}{h} \\
    = \sum_{l,h}\omega(lh^{-1},h)\braket{\psi}{\psi_l}\ketbra{hl^{-1}}{h}  
    =  \sum_{l,h}\omega^*(l,h^{-1})\braket{\psi}{\psi_l}\ketbra{hl^{-1}}{h} 
    = \sum_{l}\braket{\psi}{\psi_l}R_l^*,
\end{multline}
where the second-to-last equality follows from Eq.~(\ref{multiplier associativity cond}) and assumption~(\ref{assump gg-1}). 
%
%
Thus, in a manner analogous to the linear case, the Gram matrix \( G \) is seen to belong to the associative algebra spanned by \( \{R^*_g\}_{g\in\mathcal{G}} \). Consequently, any function of \( G \) must also lie within this algebra. In particular, its square root \( S = \sqrt{G} \) takes the form  
\begin{equation}
    S = \sum_{g\in\mathcal{G}} s_g R_g^*
\end{equation}
where \( \{s_g\}_{g\in\mathcal{G}} \) are complex coefficients.

Consider now the ensemble \( \mathcal{E}_S = \{\ket{\phi_g}, g \in \mathcal{G}\} \) constituted by the columns of \( S \), i.e., \( \ket{\phi_g} = \sum_{h \in \mathcal{G}} S_{h,g} \ket{h} \). This ensemble is also group-generated by \( \mathcal{G} \) and transforms under the projective left-regular representation, as shown by the following:
\begin{equation}
    L_g \ket{\phi_h} = \sum_{l \in \mathcal{G}} S_{l,h} L_g \ket{l} = \sum_{l \in \mathcal{G}} \omega(g,l) S_{l,h} \ket{gl} = \sum_{l \in \mathcal{G}} \omega(g,g^{-1}l) S_{g^{-1}l,h} \ket{l} = \omega(g,h) \ket{\phi_{gh}},
\end{equation}
where the identity \( \omega(g,g^{-1}l) S_{g^{-1}l,h} = \omega(g,h) S_{l,gh} \), used in the last equality, is analogous to Eq.~(\ref{S belongs to R}) and follows from the fact that \( S \) belongs to the algebra spanned by \( \{R_g^*\}_{g \in \mathcal{G}} \).

The above discussion shows that the systematic procedure used to derive the structure of the Gram matrix from the Cayley table~\( \mathscr{R} \) in the linear case also applies to projective representations, provided that condition~(\ref{assump gg-1}) holds. As noted, for any given representation, one can always find an equivalent one —corresponding to the same physical situation— in which this condition is satisfied. Therefore, the procedure can always be applied, yielding the Gram matrix of any group-generated ensemble in a specific (canonical) form, with its structure entirely determined by the symmetry of the ensemble.

To illustrate our findings, we conclude the section with a detailed, physically motivated example of a group-generated set of four states that defines a Symmetric Informationally Complete POVM (SIC-POVM) with the symmetries of a regular tetrahedron.

\begin{adjustwidth}{2em}{0em}
{\small\rm
\noindent{\em Example.} Consider the group of 3D rotation matrices \(\mathfrak{G} = \{\openone_3, \mathcal{R}_x(\pi), \mathcal{R}_y(\pi), \mathcal{R}_z(\pi)\}\), where \(\mathcal{R}_j(\theta)\) represents a rotation by an angle \(\theta\) about the \(j\)-axis. By acting on the vector \(\boldsymbol{v}_{\rm A} = (1, 1, 1)\), the group \(\mathfrak{G}\) generates the vertices A, B, C, and D of a regular tetrahedron, with corresponding position vectors 
$\{\boldsymbol{v}_{\alpha}\}_\alpha$ (\(\alpha = \text{A}, \text{B}, \text{C}, \text{D}\)), represented in Fig.~\ref{fig:tetra}. The group \(\mathfrak{G}\) forms a subgroup of the tetrahedral group \(T_d\) —the full symmetry group of a regular tetrahedron, including both rotations and reflections. The three elements \(\mathcal{R}_j(\pi)\) for \(j = x, y, z\) correspond to 180-degree rotations about the axes passing through the midpoints of pairs of opposite edges: \((\text{AB}, \text{CD})\), \((\text{AC}, \text{BD})\), and \((\text{AD}, \text{BC})\), respectively.
%
\begin{figure}[ht]
\center
\begin{tikzpicture}[scale=1.2, line join=round, line cap=round]
    \coordinate (A) at (1,1,1);
    \coordinate (B) at (1,-1,-1);
    \coordinate (C) at (-1,1,-1);
    \coordinate (D) at (-1,-1,1);

    \draw (A) -- (B); 
    \draw (A) -- (C); 
    \draw[dashed] (A) -- (D); 
      \draw (C) -- (D); 
       \draw (B) -- (D); 
         \draw (B) -- (C); 

    \node[above right] at (A) {\footnotesize A, $(1,1,1)$};
    \node[below right] at (B) {\footnotesize B, $(1,-1,-1)$};
    \node[above left] at (C) {\footnotesize C, $(-1,1,-1)$};
    \node[below left] at (D) {\footnotesize D, $(-1,-1,1)$};
\end{tikzpicture}
\caption{Regular tetrahedron generated by \(\mathfrak{G} = \{\openone, \mathcal{R}_x(\pi), \mathcal{R}_y(\pi), \mathcal{R}_z(\pi)\}\) acting on $(1,1,1)$.}
\label{fig:tetra}
\end{figure}
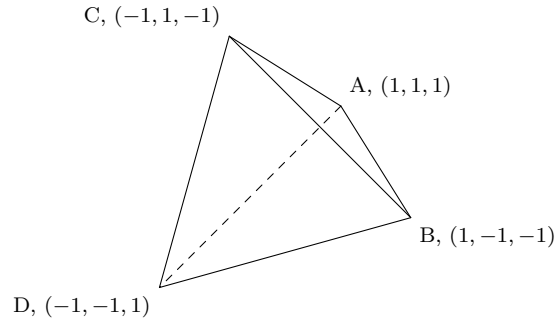

This group is isomorphic to \(\mathbb{Z}_2 \times \mathbb{Z}_2\), also known as the Klein four-group, with the following multiplication table:
\begin{center}
    \begin{tblr}{
  vline{2} = {-}{},
  hline{2} = {-}{},
  column{1} = {c}, column{2} = {c}, column{3} = {c}, column{4} = {c}, column{5} = {c}
}
  \(\mathbb{Z}_2 \times \mathbb{Z}_2\) & \(e\)   & \(x\) & \(y\) & \(z\)  \\
  \(e\)   & \(e\)   & \(x\) & \(y\) & \(z\)\\
  \(x\) & \(x\) & \(e\)   & \(z\) & \(y\)\\
  \(y\) & \(y\) & \(z\) & \(e\)   & \(x\) \\
  \(z\) & \(z\) & \(y\) & \(x\) & \(e\)
\end{tblr}
\end{center}
Explicitly, the isomorphism between the two groups is given by \(\openone \mapsto e\) and \(\mathcal{R}_j(\pi) \mapsto j\), for \(j = x, y, z\). To avoid unnecessary proliferation of symbols, we will use the same letters introduced in this table to denote the rotations themselves, as well as the elements of other equivalent representations that will be introduced below.

The family of spin-\(\frac{1}{2}\) states, \(\{|\boldsymbol{n}_\alpha\rangle \in \mathbb{C}^2\}_\alpha\), where \(\boldsymbol{n}_\alpha = \boldsymbol{v}_\alpha / |\boldsymbol{v}_\alpha|\), consists of four qubits whose Bloch vectors point towards the vertices of a regular tetrahedron. These states are important in quantum information theory, as they define a SIC-POVM~\cite{sic-povm} for qubits, with the corresponding POVM elements given by \(\{E_\alpha = (1/2) |\boldsymbol{n}_\alpha\rangle \langle \boldsymbol{n}_\alpha|\}_\alpha\).

The family also defines an ensemble of pure states generated by the action of the group \(\mathfrak{G}\), given by \(\mathcal{E} = \{ U_g |\boldsymbol{n}_{\rm A}\rangle : g \in \mathscr{G} \}\), where \(U_g \in SU(2)\) represents the standard spin-\(\frac{1}{2}\) representation of the 3D rotations in \(\mathfrak{G}\). Specifically, 
\begin{equation}
U_e = \openone_2, \quad U_j = -i \sigma_j, \quad j = x, y, z,
\end{equation}
with \(\sigma_j\), \(j=x, y, z\), being the Pauli matrices. In terms of the standard \(\mathbb{C}^2\) (qubit) basis, the seed state is given by
\begin{equation}
|\boldsymbol{n}_{\rm A}\rangle = a_- |0\rangle + e^{i\pi/4} a_+ |1\rangle,
\end{equation}
where \(a_\pm = \left[ (3 \pm \sqrt{3})/6 \right]^{1/2}\).

The set \( \{\openone_2, -i\sigma_x, -i\sigma_y, -i\sigma_z\} \), more conveniently written as \( \{e, x, y, z\} \), where
\begin{equation}
e := \openone_2, \quad j := -i\sigma_j, \quad j = x, y, z,
\end{equation}
constitutes a projective representation of the group \( \mathfrak{G} \) on \( \mathbb{C}^2 \). The corresponding multiplication table, \( \mathscr{R} \), as defined in Eq.~(\ref{formal Cayley}), is given by
\begin{center}
    \begin{tblr}{
  vline{2} = {-}{},
  hline{2} = {-}{},
  column{1} = {c}, column{2} = {r}, column{3} = {r}, column{4} = {r}, column{5} = {r}
}
  \( \mathbb{Z}_2 \times \mathbb{Z}_2 \)  & \( e \)   & \( x \) & \( y \) & \( z \)  \\
  \( e \)   & \( e \)   & \( x \) & \( y \) & \( z \)\\
  \( x \) & \( x \) & \( -e \)   & \( z \) & \( -y \)\\
  \( y \) & \( y \) & \( -z \) & \( -e \)   & \( x \) \\
  \( z \) & \( z \) & \( y \) & \( -x \) & \( -e \)
\end{tblr}
\end{center}
This table clearly reveals that \( \omega(g, g^{-1}) = -1 \) for \( g \neq e \), and thus condition~(\ref{assump gg-1}) does not hold in this representation. 

We now define an equivalent representation through the map given in Eq.~(\ref{mu=1/sqrt w}), so that condition~(\ref{assump gg-1}) is satisfied. Choosing \( \sqrt{-1} = -i \), we obtain the mapping:
\begin{equation}
e \mapsto \openone_2, \qquad j \mapsto \frac{-i\sigma_j}{\sqrt{-1}} = \sigma_j, \quad j = x, y, z.
\end{equation}
The set \( \{\openone_2, \sigma_x, \sigma_y, \sigma_z\} \) is a well-known two-dimensional projective irrep of the group \( \mathbb{Z}_2 \times \mathbb{Z}_2 \). This representation, often referred to as the Pauli representation, is closely related to the quaternion group, as the Pauli matrices satisfy multiplication rules analogous to those of the quaternions.

In the example at hand, the use of the Pauli representation is equivalent to assigning phase factors to three of the original states in the ensemble \( \mathcal{E} \). Specifically, it replaces the states \( |\boldsymbol{n}_{\rm B}\rangle \), \( |\boldsymbol{n}_{\rm C}\rangle \), and \( |\boldsymbol{n}_{\rm D}\rangle \) from the original ensemble by the physically equivalent states \( i |\boldsymbol{n}_{\rm B}\rangle \), \( i |\boldsymbol{n}_{\rm C}\rangle \), and \( i |\boldsymbol{n}_{\rm D}\rangle \), respectively.

Denoting \( \{\openone_2, \sigma_x, \sigma_y, \sigma_z\} \) by the symbols \( e \), \( x \), \( y \), and \( z \) for conciseness, the table \( \mathscr{R} \) of the Pauli representation is given~by
\begin{center}
    \begin{tblr}{
  vline{2} = {-}{},
  hline{2} = {-}{},
  column{1} = {c}, column{2} = {r}, column{3} = {r}, column{4} = {r}, column{5} = {r}
}
 \(\mathbb{Z}_2 \times \mathbb{Z}_2\)   & \(e\)   & \(x\) & \(y\) & \(z\)  \\
  \(e\)   & \(e\)   & \(x\) & \(y\) & \(z\)\\
  \(x\) & \(x\) & \(e\)   & \(i z\) & \(-iy\)\\
  \(y\) & \(y\) & \(-iz\) & \(e\)   & \(ix\) \\
  \(z\) & \(z\) & \(iy\) & \(-ix\) & \(e\)
\end{tblr}
\end{center}
From this table, we immediately obtain the canonical Gram matrix structure of any ensemble generated by any projective representation equivalent to the Pauli matrices:
\begin{equation}
    G = 
  \begin{pmatrix}
      1 & \,\,c_x & \,\,c_y & \,\,c_z \\[3pt]
    c_x & \,\,1 & \,\,ic_z & \,\,-ic_y \\[3pt]
    c_y & \,\,-ic_z & \,\,1 & \,\,ic_x \\[3pt]
    c_z & \,\,ic_y & \,\,-ic_x & \,\,1 
    \end{pmatrix},\qquad c_j\in\mathbb{R},\;  j=x,y,z,\quad G\ge0.
\end{equation}

For the SIC-POVM set of equiprobable pure states \( \{|\boldsymbol{n}_{\rm A}\rangle,\, i |\boldsymbol{n}_{\rm B}\rangle,\, i |\boldsymbol{n}_{\rm C}\rangle,\, i |\boldsymbol{n}_{\rm D}\rangle \} \), the Gram matrix takes the specified form with \( c_x = c_y = c_z = 1/\sqrt{3} \). A different choice of seed state would result in different values for the real coefficients \( c_j \), but the overall structure of the Gram matrix is determined solely by the symmetry group and the projective representation that generates the ensemble.


For completeness, we provide the matrices of the right-regular representation, derived from the last table as discussed in~Sec.~\ref{proj repr}:

\begin{equation}
R_e = \openone_4,\quad R_x = \begin{pmatrix} 
0 & 1 & 0 & 0 \\
1 & 0 & 0 & 0 \\
0 & 0 & 0 & -i \\
0 & 0 & i & 0
\end{pmatrix},
\quad R_y = \begin{pmatrix}
0 & 0 & 1 & 0 \\
0 & 0 & 0 & i \\
1 & 0 & 0 & 0 \\
0 & -i & 0 & 0
\end{pmatrix},
\quad R_z = \begin{pmatrix}
0 & 0 & 0 & 1 \\
0 & 0 & -i & 0 \\
0 & i & 0 & 0 \\
1 & 0 & 0 & 0
\end{pmatrix}.
\end{equation}
This representation decomposes into two copies of the irreducible 2-dimensional Pauli representation. The unitary matrix that achieves this block diagonalization of the right-regular representation is
%
\begin{equation}
\mathscr{U} = -\frac{i}{2} \begin{pmatrix}
1 & -1 & i & 1 \\
-1 & 1 & i & 1 \\
-1 & -1 & i & -1 \\
-1 & -1 & -i & 1
\end{pmatrix}.
\end{equation}
More explicitly,
\begin{equation}
\mathscr{U} R_j \mathscr{U}^\dagger = \sigma_j \oplus \sigma_j, \quad j = x, y, z.
\end{equation}
Since this irrep has dimension 2, the results from the main text imply that the maximum eigenvalue of \( G \) is doubly degenerate, ensuring that perfect exclusion is possible for the ensemble \( \mathcal{E} \).
}
\end{adjustwidth}

\section{Optimality of minimum-error exclusion measurement}

To prove the optimality of the POVM in Eq.(\ref{eq:non_perf_ansatz}) for \mbox{minimum-error} QSE \mbox{($\gamma=1$)} outside the region of perfect exclusion, we need to check that the operators \mbox{$\rho_g/|\mathcal{G}|-Y$}, introduced in Eq.~(\ref{eq:min_error_dual_sdp}), with \mbox{\( Y = (1/|{\cal G}|) \sum_{g\in{\cal G}} U_g \ket{\omega} \braket{\omega}{\psi} \bra{\psi} U^\dagger_g \)}, are positive semidefinite for all $g\in\mathcal{G}$. Due to group covariance, it suffices to check that positive semidefiniteness holds for $|\psi\rangle\langle\psi|-|\mathcal{G}|Y$, since
\begin{equation}
U_g\left(|\psi\rangle\langle\psi|-|\mathcal{G}|Y\right)U^\dagger_g=|\mathcal{G}|\left({|\psi_g\rangle\langle\psi_g|\over |\mathcal{G}|}-Y\right)=|\mathcal{G}|\left({\rho_g\over |\mathcal{G}|}-Y\right)\!.
\end{equation}

To do so, we define 
\begin{equation}
\label{some defs}
\beta_k^\mu:=\sqrt{\alpha_k^\mu\over |\mathcal{G}|};\quad
    \zeta:=\sqrt{|\mathcal{G}|}\,\langle\psi|\omega\rangle=\beta_1^1-\sum_{\mu\neq 1}\sum_{k=1}^{r_\mu}d_\mu\beta_k^\mu;\quad
    \ket{\xi_k^\mu}:=\ket{v_k^\mu}\ket{u_k^\mu};
\end{equation}
where $\{|v^\mu_k\rangle\}_{k=1}^{d_\mu}$, $\{|u^\mu_k\rangle\}_{k=1}^{m_\mu}$ are the (Schmidt) bases introduced in Eq.~(\ref{bipartite_decomp}). Recall that the irrep $\mu=1$ corresponds to the largest eigenvalue of the ensemble operator $\Omega$, see Eq.~(\ref{Omega}), and necessarily has dimension $d_1=1$ in the region considered, so $\zeta > 0$.
By the Great Orthogonality Theorem, Eq.(\ref{eq_app:great_ortho_relation}), the operator $|\mathcal{G}|Y$ can be express as
\begin{equation}
\label{messy 1}
        |\mathcal{G}|Y =  
         \left(\beta^1_1\right)^2\ketbra{\xi^1_1}{\xi^1_1}-\\
         \zeta\sum_{\mu\neq1}\sum_{k=1}^{r_\mu}\beta_k^\mu\Big(d_\mu\ketbra{\xi^1_1}{\xi^1_1} + \ketbra{\xi_k^\mu}{\xi_k^\mu}\Big)-
 \Bigg(\sum_{\mu\not=1}\sum_{k=1}^{r_\mu}d_\mu\beta^\mu_k\Bigg)^2\ketbra{\xi^1_1}{\xi^1_1}-\zeta Z,
\end{equation}
where $Z$ is the positive semidefinite operator
\begin{equation}
    Z = \sum_{\mu\neq1}\sum_{j=1}^{r_\mu}\sum_{\substack{k=1\\ k\neq j}}^{d_\mu}\beta_j^{\mu}\ketbra{v_k^\mu}{v_k^\mu}\otimes\ketbra{u_j^\mu}{u_j^\mu}.
\end{equation}
The seed state of the ensemble is $|\psi\rangle=\sum_\mu \sqrt{d_\mu} (|\psi_\mu\rangle/\sqrt{|\mathcal G|})$, and from Eq.~(\ref{bipartite_decomp}), $\ket{\psi_\mu}/\sqrt{|\mathcal{G}|} =\sum_{k=1}^{r_\mu}\beta_k^\mu\ket{\xi_k^\mu}$. Using this and Eq.~(\ref{messy 1}) we obtain
\begin{equation}\label{eq_app:M_operator_decomp}
  |\psi\rangle\langle\psi| -|\mathcal{G}|Y = \zeta Z+ \zeta\sum_{\mu\neq1}\sum_{k=1}^{r_\mu}\beta_k^\mu\ketbra{\phi_k^\mu}{\phi_k^\mu} + \ketbra{\Psi}{\Psi},
\end{equation}
where
\begin{equation}
    \ket{\phi_k^\mu} := \sqrt{d_\mu}\ket{\xi^1_1}+\ket{\xi_k^\mu};\quad
    \ket{\Psi} := \sum_{\mu\neq1}\sum_{k=1}^{r_\mu}\sqrt{d_\mu}\beta_k^\mu\ket{\phi_k^\mu}.
\end{equation}
%
Since each term on the right-hand side of Eq.~(\ref{eq_app:M_operator_decomp}) is positive semidefinite, it follows that \( |\psi\rangle\langle\psi| - |\mathcal{G}|Y \geq 0 \),  
proving the optimality of the POVM generated by the seed state \( \ket{\omega} \) in Eq.~(\ref{eq:non_perf_ansatz}). \qed
%


\section{Optimality of unambiguous exclusion POVM}

In this section, we prove that the POVM seed state given in Eqs.~(\ref{eq:non_perf_ansatz}) and (\ref{gamma unamb}) is optimal, i.e., it minimizes the failure probability \( Q \). 

The primal SDP in Eq.~(\ref{eq:zero_error_sdp}) specializes to the group-covariant scenario at hand as  
\begin{equation}
         \min_{E\ge0}\;1-\Tr(\Omega E),\quad
         \text{subject to}\;  \openone-\sum_{g\in\mathcal{G}}U_gEU_g^\dag\geq0,\;
       \bra{\psi}E\ket{\psi}=0,
\end{equation}
where \(E\) is the POVM seed operator, satisfying \(\Pi_g = U_g E U^\dagger_g\) for all \( g \in \mathcal{G} \), with no assumption on the rank of \( E \) at this stage.  
The corresponding dual SDP is given in Eq.~(\ref{eq:dual_sdp_UA}):  
\begin{equation}\label{eq:dual_sdp_UA_suppl}  
        \max_{X} \;1-\Tr(X), \quad \text{subject to} \;
        \sum_{g\in\mathcal{G}}U_gXU_g^\dag + \nu\ketbra{\psi}{\psi}-\Omega \geq 0,\;  
        \nu\in\mathbb{R}.  
\end{equation}

It can be verified that the dual problem satisfies the Slater conditions~\cite{boyd_2004,watrous_2018}, i.e., the constraint set in Eq.~(\ref{eq:dual_sdp_UA_suppl}) has a strictly feasible point, ensuring strong duality. In other words, the primal and dual problem yield the same objective value. However, the primal problem lacks a strictly feasible solution due to the rank-deficient constraint on~$E$, creating an asymmetry in the Slater conditions between the two programs.
As noted in~\cite{watrous_2018}, such asymmetry can result in situations where the solution to one problem may not be strictly attainable. This is the case here, and complementary slackness cannot be used to derive the dual solution from the primal one. 

To address this, we propose an ansatz for \( X \) that provides a lower bound on the optimal inconclusive probability. We show below that this lower bound matches the upper bound \( Q^{\text{prim}} \) from Eq.~(\ref{eq:Pe-unambiguous}) in a specific limit. Our ansatz is  
\begin{equation}
    X =  \bigoplus_{\mu\neq1}\openone_{d_\mu}\otimes\sum_{k=1}^{r_\mu}\left[\beta_k^\mu\left(\beta_k^\mu+\Delta\right)+\frac{\varepsilon}{d_{\Lambda}}\delta_{\mu,\Lambda}\delta_{k,r_\Lambda}\right]\ketbra{u_k^\mu}{u_k^\mu},
\end{equation}
where \( \Delta := \sum_{\mu\neq1}\sum_{k=1}^{r_\mu} d_\mu \beta_k^\mu \), \( \varepsilon \) is an arbitrary positive constant, and the coefficients \( \beta^\mu_k \) are defined in Eq.~(\ref{some defs}).  
We adhere to the notation of the main text and order the eigenvalues of the ensemble operator \( \Omega \) (and the Gram matrix~\( G \)) in decreasing order, from largest to smallest. Following this convention, the indices \( \mu = \Lambda \), \( k = r_\Lambda \) correspond to (any of) the smallest nonzero eigenvalue(s) of \( \Omega \). 
We obtain  
\begin{equation}
1-\Tr(X) = 1 - \sum_{\mu\neq1} d_\mu \sum_{k=1}^{r_\mu} \beta^\mu_k (\beta^\mu_k + \Delta) - \varepsilon  
= (\beta^1_1)^2 - \Delta^2 - \varepsilon = Q^{\text{prim}} - \varepsilon ,
\end{equation}
where we have used the normalization condition of the seed state \( |\psi\rangle \):  
\begin{equation}
1 = \sum_{\mu} d_\mu \sum_{k=1}^{r_\mu} (\beta^\mu_k)^2.
\end{equation}

We next prove that for any \( \varepsilon > 0 \), \( X \) is feasible. Since \( X \) is group-invariant, the condition in Eq.~(\ref{eq:dual_sdp_UA_suppl}) simplifies to  
\begin{equation}\label{X feasible}
        X + \nu |\psi\rangle\langle\psi| - \frac{\Omega}{|\mathcal{G}|} \geq 0,
\end{equation}
where we have rescaled \( \nu \) to absorb the constant factor \( |\mathcal{G}| \). We need to show that for any \( \varepsilon > 0 \), there exists \( \nu \) such that Eq.~(\ref{X feasible}) holds. Since we are interested in the limit \( \varepsilon \to 0^+ \) and expect \( \nu = \mathcal{O}(\varepsilon^{-1}) \) in this limit, we write \( \nu = \tilde{\nu} / \varepsilon \), where \( \tilde{\nu} \) is a positive constant (of order one).  
Using some of the definitions introduced in Eq.~(\ref{some defs}), we can write  
\begin{equation}\label{feasib X decomp}
X + \frac{\tilde{\nu}}{\varepsilon} |\psi\rangle \langle \psi| - \frac{\Omega}{|\mathcal{G}|} = \Delta Z + \frac{\varepsilon}{d_\Lambda} \left(\; \sum_{\substack{k=1 \\ k \neq r_\Lambda}}^{d_\Lambda} |v^\Lambda_k\rangle \langle v^\Lambda_k| \right) \otimes |u^\Lambda_{r_\Lambda}\rangle \langle u^\Lambda_{r_\Lambda}| + K ,
\end{equation}
where  
\begin{equation}\label{K-1}
        K := \frac{\tilde{\nu}}{\varepsilon} \ketbra{\psi}{\psi} - (\beta^1_1)^2 \ketbra{\xi^1_1}{\xi^1_1} + \sum_{\mu \neq 1} \sum_{k=1}^{r_\mu} \left( \beta_k^\mu \Delta + \frac{\varepsilon}{d_{\Lambda}} \delta_{\mu,\Lambda} \delta_{k,r_\Lambda} \right) \ketbra{\xi_k^\mu}{\xi_k^\mu}.
\end{equation}
Since the first two operators on the right-hand side of Eq.~(\ref{feasib X decomp}) are manifestly positive semidefinite, the feasibility of~\( X \) reduces to proving the positive definiteness of \( K \), which we now proceed to do.

To obtain clearer expressions, we introduce a new notation: we assign a single integer index to replace each pair~\( (\mu, k) \), and write, e.g., \( \beta_p \) and \( |p\rangle \) instead of \( \beta^\mu_k \) and \( |\xi^\mu_k\rangle \), respectively. The integer indices are assigned such that \( \beta_1 \, (= \beta^1_1) > \beta_2 \ge \beta_3 \ge \cdots \ge \beta_N \, (= \beta^\Lambda_{r_\Lambda}) > 0 \). In this notation, Eq.~(\ref{K-1}) becomes:  
\begin{equation}
        K = \frac{\tilde{\nu}}{\varepsilon} \sum_{p,q=1}^N \sqrt{d_p d_q} \, \beta_p \beta_q \ketbra{p}{q} - (\beta_1)^2 \ketbra{1}{1} + \sum_{p > 1}^N \left( \beta_p \sum_{q > 1}^N d_q \beta_q + \frac{\varepsilon}{d_N} \delta_{p,N} \right) \ketbra{p}{p},
\end{equation}
from which the matrix of \( K \) is easily obtained. This allows us to compute its pivots through row elimination and, in turn, its leading principal minors. Denoting these minors by \( \det K_l \), where \( l \) is the order of the corresponding submatrix, we obtain the following:  
\begin{align}
    & \det K_1 = \frac{\tilde{\nu}}{\varepsilon} \beta_1^2 + \mathcal{O}(1), \\
    & \det K_l = \frac{\tilde{\nu}}{\varepsilon} \beta_1^2 \left(\; \prod_{p=2}^l \beta_p \right) \left(\; \sum_{p=l+1}^N d_p \beta_p \right) \left( \;\sum_{p=2}^N d_p \beta_p \right)^{\!\!l-2} \!+\; \mathcal{O}(1), \quad 2 \leq l \leq N-1, \\
    & \det K_N = \tilde{\nu} \beta_1^2 \left(\; \prod_{p=2}^N \beta_p \right) \left(\; \sum_{p=2}^N d_p \beta_p \right)^{\!\!N-3} \!\!+\; \mathcal{O}(1),
\end{align}
where the terms that depend on \( \tilde{\nu} \) are explicitly shown. Thus, for any \( \varepsilon > 0 \), and independently of the \( \mathcal{O}(1) \) terms, these minors can be made strictly positive by choosing \( \tilde{\nu} \) sufficiently large. Hence, by Sylvester's criterion~\cite{horn_2012}, \( K \) is positive definite for this choice of \( \tilde{\nu} \).


In summary, we have shown that for any \( \varepsilon > 0 \), the following inequality holds:
\begin{equation}
    Q^{\text{prim}} - \varepsilon \leq Q^* \leq Q^{\text{prim}},
\end{equation}
where \( Q^* \) denotes the optimal probability of obtaining an inconclusive outcome. Taking the limit \( \varepsilon \to 0^+ \), we obtain the final result: \( Q^* = Q^{\text{prim}} = Q^{\rm min} \).
\qed

\end{document}